\documentclass[1p]{elsarticle}

\usepackage[english]{babel}

\usepackage{amsmath}
\usepackage{graphicx}
\usepackage[colorlinks=true, allcolors=blue]{hyperref}

\title{Ethics through the Facets of Artificial Intelligence \tnoteref{t1}}
\tnotetext[t1]{This work received partial financial support from FAPESP -- Project 2020/09850-0.}
\author[1]{Flavio S. Correa da Silva}
\affiliation[1]{
    organization = {Institute of Advanced Studies -- University of Sao Paulo},
    addressline = {Rua da Praca do Relogio 109},
    postcode = {05508-090 email: fcs@usp.br},
    city = {Sao Paulo},
    country = {Brazil}
    }

\begin{document}

\begin{abstract}

Artificial Intelligence (AI) has received unprecedented attention in recent years, raising ethical concerns about the development and use of AI technology. In the present article, we advocate that these concerns stem from a blurred understanding of AI, how it can be used, and how it has been interpreted in society. We explore the concept of AI based on three descriptive facets and consider ethical issues related to each facet. Finally, we propose a framework for the ethical assessment of the use of AI.
\end{abstract}


\maketitle

\section{Introduction}

The name \textit{Artificial Intelligence (AI)} denotes a multifaceted field of human knowledge and inquiry. In the present article, we consider three of the multiple facets of AI and how these facets are connected. Each facet has been given prominence at different moments in the brief history of AI:

\begin{enumerate}
    \item \textbf{AI as a scientific endeavour}: AI was initially proposed as a topic for scientific research, following a rigorous and goal-oriented research methodology. AI as a theme for rigorous scientific inquiry, which was predominant at a given time, has persisted to this day, although discreetly, given that it now needs to share the attention with other facets of AI.
    \item \textbf{AI as a technology}: AI research has required the development of specialised conceptual and computational tools for the design and execution of experiments. These tools have proven to be effective for problem solving in general, in many cases detached from the initial purposes for which they have been developed. As a consequence, the interpretation of AI now encompasses the development of tools which are capable of solving problems and have been inspired by the observation of intelligent behaviour.
    \item \textbf{AI as a sociotechnical imaginary}: the results obtained in AI, either as a scientific endeavour or as an inspiration to build tools for problem solving, have been opportunistically adopted as raw material to fabricate social, cultural, and economic scenarios, employed to steer strategic decisions towards directions which are of economic and political interest of specific groups of influence.
\end{enumerate}

These facets influence each other, but mutual influences have not always been constructive. In the present article, we introduce a perspective of the field of AI based on these three facets, which we believe can be useful to expose how and why AI is moving in the direction and at the pace it is, and to suggest alternative directions which may be more beneficial and ethically reliable for society than what we see nowadays.  

In Sections \ref{science}, \ref{technology} and \ref{imaginary} we discuss respectively the \textit{scientific}, \textit{technological} and \textit{sociotechnical imaginary} facets of AI. In Section \ref{ethics} we consider ethical implications of the connections across these facets. In Section \ref{conclusion} we draw some conclusions and final considerations.

\section{AI as a scientific endeavour}
\label{science}

AI as a field of human knowledge and inquiry presents the unusual feature of having a ``birth certificate", given that the name \textit{Artificial Intelligence} was first used to characterise a novel research field at a workshop held at Dartmouth College (USA) in 1956. The proposal to organise the Dartmouth Workshop aimed at the validation of the conjecture that \textit{``(...) every aspect of learning or any other feature of intelligence can in principle be so precisely described that a machine can be made to simulate it"} \cite{McCarthy_Minsky_Rochester_Shannon}. 
It focused, therefore, on \textit{observable intelligent behaviour} instead of intelligence \textit{per se}, presumably considering that rigorous scientific scrutiny could be better applied to observable features of a phenomenon, rather than the phenomenon itself.

Furthermore, the proposed research initiative focused on \textit{computationally executable models}, which can be viewed as a computer-based rendition of a predicament formulated by Leonardo da Vinci around 450 years earlier:

\begin{quote}
    \textit{''No human investigation can be termed true science if it is not capable of mathematical demonstration. If you say that the sciences which begin and end in the mind are true, I do not agree, but deny it for many reasons, and foremost among these is the fact that the test of experiment is absent from these exercises of the mind, and without these there is no assurance of certainty"} (as quoted in \cite{isaacson2019leonardo}).
\end{quote}

AI as a scientific endeavour, therefore, can be summarised from the Dartmouth Workshop proposal as
\begin{quote}
    \textit{the study of observable features of intelligent behaviour, through rigorous mathematical modelling and implementation as executable computational models to be used for empirical validation with respect to their biological counterparts.} 
\end{quote}
It should be noticed that a scientific programme formulated this way precludes concepts such as ''General Intelligence" -- referring to a model of intelligence that could be considered fully equivalent to the biological entities it should be modelling -- or ''Super Intelligence" -- referring to a model of intelligence that could somehow surpass the biological entities it should be modelling, and hence would invalidate the experiments considering their \textit{raison d'être}, namely behavioural alignment with respect to the biological references.

The workshop proposal structured the newborn field of AI along seven axes:

\begin{enumerate}
    \item \textbf{Automatic computers}, aiming at the development of computationally efficient systems that encode generic problem solving methods that mimic human deductive problem solving. This axis evolved later to the field of \textit{Knowledge Representation and Inference}.
    \item \textbf{''How can a computer be programmed to use a language?"}, aiming at the development of Computational Linguistics and \textit{Natural Language Processing}.
    \item \textbf{Neuron nets}, aiming to study the morphology and physiology of the brain as an information processing device. 
    \item \textbf{''Theory of the size of a calculation"}, aiming at computational theorem proving and methods for automating mathematical proof synthesis. This axis evolved to the field of \textit{Automated Reasoning}.
    \item \textbf{Self improvement}, aiming at the development of systems capable of adaptation and evolution given environmental stimuli. This axis evolved to the field of \textit{Evolutionary Computing} and, in combination with \textbf{neuron nets}, the field of incremental function approximations, popularised under the name \textit{Machine Learning}.
    \item \textbf{Abstractions}, aiming at the formal characterisation and computational implementation of systems to build conceptual abstractions given observations. This axis evolved to the field of \textit{Ontological Reasoning}.
    \item \textbf{Randomness and creativity}, based on the assumption that creativity could be simulated using controlled bias of random variables.
\end{enumerate}

Fifty years after the Dartmouth Workshop, a second event was held at Dartmouth College, with the participation of important researchers in AI and focusing on past achievements and future perspectives for the field \cite{moor2006dartmouth}. Optimistic, forward-looking views of AI at that moment consolidated the methodology adopted for the research field as a whole, as well as the seven axes used as guidelines to organise the field, although revised and further branched out, yet preserving the originally proposed general structure.

Some critical observations, however, can be made after careful reading of Moor's account of the second Dartmouth Workshop that occurred in 2006 \cite{moor2006dartmouth}:

\begin{itemize}
    \item As observed by Marvin Minsky, who was a coauthor of the first workshop proposal and participant in the second workshop, \textit{''(...) too many in AI today try to do what is popular and publish only successes (...) AI can never be a science until it publishes what fails as well as what succeeds"} (as quoted in \cite{moor2006dartmouth}).

    \item The AI research community was split between the then called \textit{neats} and \textit{scruffies} \cite{poirier2024neat}. More significantly than the split itself, these two groups were characterised in terms of \textit{how} theories were implemented by each group for empirical validation (logics and automated proof systems for \textit{neats} versus stepwise approximations of functions computationally represented as layered networks of simplified units for \textit{scruffies}) instead of \textit{why} these theories were being developed and what could be covered by models built according to each group (foundational explanatory models for \textit{neats} versus behavioural black box models for \textit{scruffies}). It is worth observing that the choice of techniques based on logics versus numerical function approximations does not necessarily correlate with the choice of approaches based on foundational versus behavioural modelling.
    \item Little emphasis was placed on ''cross-over research" combining different axes, e.g. composition of automated/formal reasoning and machine learning for the development of formal learning theories \cite{valiant1984theory,valiant1999robust}, composition of agent coordination and automated/formal reasoning \cite{halpern1990knowledge,halpern1992guide}, composition of agent coordination, automated/formal reasoning and machine learning \cite{blum2017collaborative,cullina2018pac}, or ambitious compositions of all observable features of intelligence \cite{laird2019soar,kotseruba202040}. Combinations of axes have proven to be enlightening for the disclosure of relevant features of intelligence that can fall beyond a simple linear composition of features that can be captured by each axis independently.
\end{itemize}

These observations indicated an already ongoing trend within the AI research community toward fragmentation in specialisations, as well as a focus on technical developments rather than contributions to the ''conceptual big picture" of AI. This trend has persisted until the present day. Despite the inevitable deceleration of scientific advancements in AI, one positive side effect of this trend has been the development of specialised technologies, as discussed in the next section.

\section{AI as a technology}
\label{technology}

As pointed out in the reflections about the second Dartmouth Workshop, AI \textit{circa} 2006 had turned towards axis-specific results and focused on the development of computational tools for experimentations to build and validate conjectures about the nature of intelligence.

The computational tools obtained have been shown to be effective for problem solving in general, independently of contributions to the field of AI \textit{per se}, and have attracted the interest of a variety of fields such as engineering, product design, and business development. As a consequence, AI started to denote not only a field of scientific endeavour, but also the development of computational tools and techniques for problem solving, inspired by results obtained from the scientific approach to AI.

Each axis for the development of AI generated specific computational tools, which can be retrospectively organised into four groups as follows:
\begin{enumerate}
    \item \textbf{Tools for computational logical reasoning and deductive problem solving}, aligned with axes 1 (automatic computers), 4 (''theory of the size of a calculation") and 6 (abstractions): prolific research was developed in the development of specialised programming languages to support the \textit{declarative programming} paradigm, most remarkably logic programming and PROLOG \cite{apt1991logic,apt1997logic}; automated reasoning for non-classical logics applied to knowledge representation and inference \cite{fagin2004reasoning,van2008handbook}; heuristic methods for efficient solution of the boolean satisfiability problem \cite{moskewicz2001chaff,een2003extensible}; computational ontologies \cite{uschold1996ontologies,fensel2001ontologies}; and applications in knowledge representation and problem solving in engineering, business, medical reasoning and support to scientific reasoning.
    
    \item \textbf{Tools for Natural Language Processing (NLP)}, aligned with axis 2 (''how can a computer be programmed to use a language?"): the research work and results in NLP have been grounded on two approaches, which can be coined \textit{foundational} and \textit{statistical} NLP or, alternatively, \textit{neat} and \textit{scruffy} NLP. Foundational research focused on understanding language as a phenomenon and working on computational reconstructions of it, aiming to combine the goals of understanding the very nature of human language and generating useful tools for tasks such as \textit{text translation, document summarisation, information extraction} and \textit{question answering} \cite{wilks2005history}. Gradually, foundational research in NLP has given room for statistical research, which places little emphasis on understanding human language and focuses on task-oriented results based on observable self-referential co-occurrences of language patterns in discourse \cite{johri2021natural}.
    
    \item \textbf{Tools for adaptive and heuristic optimisation}, aligned with axis 5 (self improvement): research in collective intelligence and how individual behaviour can be locally and dynamically adapted to optimise performance at population level, for example, for preservation of species under hostile environmental circumstances, has led to development of techniques for local adaptive behaviour and piecewise, coordinated optimisation employing strategies that maximise probability of convergence towards global optima. The techniques obtained have proven to be useful for problem solving in complex and dynamic scenarios, characterising the subfield denoted as \textit{Evolutionary Computing} \cite{eiben2015introduction}.
    
    \item \textbf{Tools for computationally efficient stepwise function approximations}, \textit{aka} Machine Learning (ML), aligned with axes 3 (neuron nets) and 5 (self improvement): ML has received unprecedented attention in recent years, partially for reasons to be discussed in the next section, and partly due to interesting recent empirical results, inaugurated by the achievement obtained in the ImageNet Competition in 2012 \cite{krizhevsky2012imagenet}. The ML community has gradually split in two groups: one group has steered towards foundational and theoretical results, focusing either on how to model learning \textit{per se} -- hence interacting with the field of Cognitive Science \cite{perconti2020deep} -- or how to build computationally effective learning models, in the sense of being sufficiently expressive to approximate general function classes efficiently \cite{gavranovic2024thesis,jia2024category,augustine2024survey} as well as being probabilistically reliable given accepted probabilistic assumptions about the domain of interest and the available samples to characterise it \cite{valiant1984theory,valiant1999robust,levine2018deep,de2023makes}; the other group steered towards performance in benchmarks and how to best design benchmarks to highlight features of interest in engineered systems \cite{prata2024lob,kopalidis2024advances,rubachev2024tabred,wan2024deep}.
\end{enumerate}

Axis 7 (randomness and creativity) stayed closer to foundational research in cognitive sciences. Although influential and connected to the other axes and their corresponding tools, it has not been a primary source of inspiration for the development of computational tools to be used for problem solving.

In order for any tool to be considered sufficiently trustworthy to be adopted in mission critical, risk-prone, and/or costly scenarios, quality and reliability assessment methodologies must be grounded on rigorous formal analysis, transparent descriptions, and detailed guidelines and requirements for appropriate use, including data quality requirements and statistical assumptions about input data and corresponding expected outcomes. Many issues of ethical concern related to AI stem from poor assessment and inadequate utilisation of tools, either by lack of clarity in available information or by malicious misinformation about capabilities of tools.

It should be noticed that, when AI technology is used for problem solving, quality assessment is detached from comparison with intelligence as observed in biological entities, focusing instead on effectiveness and computational efficiency in solving problems.

\section{AI as a sociotechnical imaginary}
\label{imaginary}

According to the Oxford English Dictionary, the word \textit{imaginary} as a noun originates in Latin and refers to \textit{''something imagined"}, being frequently used in plural form (\textit{imaginaries}) to refer to a collection of imagined forms associated to a specified theme. Interestingly, recent use of this word in singular form, accompanied by certain modifiers, refers to imaginaries shared by groups of people: \textit{collective imaginary} \cite{ervik2023generative}, \textit{social imaginary} \cite{taylor2004social}, \textit{techno-social imaginary} \cite{gordon2009learning}, \textit{sociotechnical imaginary} \cite{jasanoff2009containing,jasanoff2013sociotechnical}. Each of these terms refers to individual collections of imaginaries, associated to themes belonging to one category, which are shared and agreed upon by a population.

A \textit{sociotechnical imaginary} refers to shared collections of imaginaries that depict forms of social life and social order dependent on the design and fulfillment of population-specific scientific and technological visions.

AI is our category of interest in this article, and in this section we focus on the \textit{sociotechnical imaginary of AI}, meaning shared collections of imaginaries 
   \textit{(1)} 
     envisioning possible consequences of utilisation of hypothetically feasible systems based on AI (considered both as models built to study intelligence as a phenomenon, which are capable of imitating biological intelligence to some extent, and tools which have been developed to enable the construction of these models and have proven to be useful for efficient problem solving in a variety of domains),
   \textit{(2)}
     assessing these consequences with respect to appropriate shared value scales considering aspects such as cultural and social values, wealth generation and distribution, quality of life, and environmental sustainability, and
   \textit{(3)} 
     determining actions to be taken to steer the development of systems based on AI to ensure benefits and prevent harms.

Some points must be clarified to ensure an effective characterisation of a sociotechnical imaginary of AI:
\begin{itemize}
    \item What is the population of interest? 
    
    Different criteria can be adopted to circumscribe a population, such as nationality (indeed, this was the criteria considered to characterise the original concept of \textit{sociotechnical imaginary}), gender, religion, age range, formal education, wealth, or even an encompassing consideration of present and future humankind as a whole. 
    
    \item What are the hypothetically feasible systems under consideration based on AI? How are these systems conceived? How do these systems connect with existing enabling technologies and scientific results?

    The most reliable sources of information to build hypotheses about the feasibility of present and future imagined systems based on AI are scientific and technical publications. Given the elevated frequency of publications about AI and considering the high level of scholarly elaboration of many of these publications, populations at large usually access interpretations of those publications, written using lay terms and, in many cases, conveying opinions and biases that were not present in the original publications.

    Misinformed interpretations can, by themselves, mislead the understanding of actual capabilities and potentialities of existing technologies and scientific results, thus leading to false hypotheses to ground the construction of envisioned consequences of the use of systems based on AI.
    
    Misinformation about the capabilities and potentialities of existing technologies and scientific results has been amplified by a purposeful distortion of the reality about AI, promoted by an oligopoly of organisations interested in profit making through the intensified use of specific computational resources that are required by a particular family of tools developed for ML \cite{bender2021dangers,paullada2021data,gebru2024tescreal,inie2024ai,monett2024deconstructing}. 
    
    Biased hypotheses about feasible systems based on AI build unrealistic scenarios about the future considering the use of such systems, and distort decisions about collective actions to be taken in the present to either foster or prevent the considered scenarios.
    
    \item What values are considered important to be preserved? How do different values compare with each other?

    Values and value scales are individually and collectively shape shifting. As a consequence, the values ascribed to specific scenarios are also dynamic and unpredictable, and action plans should take into account these value dynamics.

    Values are also susceptible to influences, e.g. through marketing malpractices and biased misinformation \cite{lemke2024ai,vallor2024ai}. Scrutable organisational guardrails maintained by reliable institutions, possibly public administration representatives and multilateral organisations, can be an appropriate course of action to prevent biased influences on the natural evolution of shared values and value scales.
    
    \item What actions are accepted to be taken collectively by the population of interest under consideration? How can these actions be implemented?

    Actions are inevitably goal-oriented, and collective coordinated actions can only be effective if goals are publicly and globally accepted by a population. Appropriate mechanisms must be built and maintained by reliable institutions, so that public, globally accepted goals and corresponding action plans can stabilise and move forward. Stabilisation of goals and action plans depends on appropriate \textit{humanity-centred design} practices \cite{russell2020humanity,norman2023design}.    
\end{itemize}

\section{AI and ethics}
\label{ethics}

The name \textit{ethics} comes from ancient Greek, originating from \textit{ethos}, the character or set of dispositions of an individual or a population. \textit{Ethics} refers to the study of \textit{ethos}. Ethical studies are commonly structured in three approaches \cite{vallor2018introduction}:

\begin{enumerate}
    \item \textbf{Virtue ethics}, which can be characterised based on the notion of a \textit{good life} as described by Greek philosophers such as Socrates, Plato, Aristotle and the Stoics (and beautifully portrayed in literary form by the modern writer N. Kazantzakis \cite{kazantzakis2012report}): one has lived a good life if, when looking back on their actions, life itself is deemed successful. Greek philosophers leaned on the problem of how to measure success and agreed on the view that a life worth living (\textit{ie} a good life) would be a life devoted to the practice of virtue.

    The natural next problem, given this portrayal of a good life, is the characterisation of virtue. Socrates, Plato, and the Stoics characterised virtue based on four fundamental manifestations, namely \textit{wisdom}, \textit{courage}, \textit{justice} and \textit{self-control}. Aristotle added other manifestations to the list, which can be interpreted as refinements of the four fundamental manifestations of his predecessors in Greek philosophy. A similar approach to the notion of good life and characterisation of virtue through its manifestations can be found in other regions and times (although varying on the actual identification and denotation of the manifestations of virtue), such as sacred writings of Jewish, Christian, Islamic, Buddhist and Taoist traditions.

    Virtue \textit{ethics} refers to the pursuit of individual behaviour that can lead to a collective cultivation of virtue, considering a characterisation of virtue based on manifestations that are agreed upon by a population.

    \item \textbf{Duty ethics}, which is based on shared and agreed moral principles (typically registered as accredited documents containing norms, regulations, laws, etc.): duty \textit{ethics} refers to the pursuit of individual behaviour that can lead to collective obedience to moral principles.

    Historically, duty ethics features the oldest documented traces of existence in comparison with the other approaches to ethics (possibly given that it is grounded on registered accounts of moral principles), as evidenced by the Babylonian \textit{Code of Hammurabi} (c. 1800 BC) and the Egyptian \textit{Laws of Maat} (c. 1200 BC) \cite{o2024ethical}.

    \item \textbf{Consequence ethics}, which has the shortest history of the three approaches to ethics: based on the British Enlightenment, it was first explicitly articulated by Jeremy Bentham (1748–1832) and John Stuart Mill (1806–1873), suggesting that individual actions should be guided by maximisation of \textit{pleasure} and minimisation of \textit{pain}. Consequence ethics is, comparatively, more objective than virtue or duty ethics, as it does not require subjective accounts of virtue and virtuous actions, nor subjective judgements about obedience to moral principles. It requires the quantification of the \textit{consequences} of actions with respect to pleasure and pain, so that a balance can be calculated to judge the moral quality of actions.
\end{enumerate}


These approaches to ethics are complementary to each other. They all deal with judgements of individual actions with respect to collectively accepted values (respectively, virtue, moral obligations, and quantified accounts of implied levels of pleasure and pain).

Individual actions can be supported and influenced by existing technological tools. The \textit{use} of available tools can ascribe ethical values to actions, although not necessarily the tools themselves (unless a tool is designed and implemented for a specialised use -- e.g. a weapon -- in which case it inherits the ethical values of this specialised use to which it is constrained). Considering the facets highlighted in the present article, AI can generate two sorts of tools:
\begin{enumerate}
    \item As a \textbf{scientific endeavour}, AI can generate tools to mimic human behaviour and participate in social interactions \textit{as if} tools were genuinely intelligent.
    \item As a \textbf{technology}, AI can generate tools for efficient problem solving and support human activities.
\end{enumerate}

Ethical considerations apply to human designers and developers, and to human users and their actions in which tools of each sort are employed.

Tools enabled by AI are owned by organisations that employ designers and developers to build them and entice users to adopt them, based on views claimed by AI as a \textbf{sociotechnical imaginary}. Ethical considerations also apply to these organisations, which indirectly make use of these tools for profit-making \textit{instead of} problem solving or imitation of human behaviour.

A framework to facilitate alignment, in practice, with ethical principles can be based on three layers of responsibilities:

\begin{enumerate}
    \item \textbf{The baseline layer can focus on \textit{transparency} and \textit{literacy}}: system designers and developers must be responsible for ensuring transparency with respect to the tools they develop, through clear and thorough documentation of their capabilities and requirements to be used; users must be trained about the tools they wish to use, to ensure proper use and proper expectations about what each tool can deliver; and owners of tools must ensure that realistic expectations are built about the tools they commercialise.
    \item \textbf{The intermediate layer can focus on \textit{ethical values}}: accredited institutions such as governments, multilateral organisations, religious institutions, educational institutions, etc. must be responsible for the establishment of ethical principles which are valued for the population under consideration (see e.g. the \textit{UNESCO Recommendation on the Ethics of Artificial Intelligence}\footnote{\tt https://www.unesco.org/en/artificial-intelligence/recommendation-ethics?hub=32618}).
    \item \textbf{The top layer can focus on \textit{regulations} to ensure that existing systems align with ethical requirements}: accredited institutions must be responsible for the establishment of methods for proper use and quality assessment of AI-based systems according to ethical values.
\end{enumerate}

Implementation and support to a framework of this sort demands effort, and ideally should be performed by accredited institutions of the sort of the ones suggested in previous paragraphs to identify and safeguard ethical values. Considering virtue ethics as a possibility for practical implementation of a framework, a first-order logical theory can be formulated, based on which the establishment of virtue-driven ethical principles and corresponding methods to align proper use and quality assessment of AI technology can be verified.

\section{Conclusion}
\label{conclusion}

Considering AI as a multifaceted concept, we have focused on three specific facets of AI. Two of these facets (namely, \textit{AI as a scientific endeavour} and \textit{AI as a technology}) generate tools that can be used somewhat independently from the original AI-related contexts for which they were created (respectively, to mimic human intelligent behaviour for a variety of purposes and to solve problems efficiently), and the third facet (namely, \textit{AI as a sociotechnical imaginary}) considers explicitly three sorts of stakeholders that interact through the utilisation and exploitation of these tools: system \textit{designers and developers}, system \textit{users} and system \textit{owners}, i.e. organisations who own and exploit existing tools for profit making.

Ethical considerations apply to different stakeholders according to each of the three facets of AI considered in this study. The fundamental approaches to ethics are also three (\textit{virtue}, \textit{duty} and \textit{consequence ethics}).

The philosopher Albert Borgmann wrote a prescient book in which the notion of \textit{good life} is connected to three different modes of interaction between humans and information \cite{borgmann1999holding}:

\begin{enumerate}
    \item \textbf{Information \textit{about} reality}, in which information is seen as the means by which humans access and interpret reality and therefore interact with the physical world, other humans and abstract concepts such as emotions and imagined events.
    \item \textbf{Information \textit{for} reality}, in which information is seen as the means by which humans generate goals and, based on information \textit{about} reality, devise plans to modify reality.
    \item \textbf{Information \textit{as} reality}, in which information \textit{replaces} reality, in such a way that humans no longer access, interpret or act upon the physical world, other humans, and abstract concepts. 
\end{enumerate}

There is a correspondence between each facet of AI explored in this article and each mode of interaction between humans and information proposed by Borgmann:

\begin{enumerate}
    \item AI as a \textbf{scientific endeavour} generates tools to imitate human intelligent behaviour, this way portraying the reality \textit{about} aspects of human intelligence.
    \item AI as a \textbf{technology} generates tools for efficient problem solving, with the aim of prescribing actions to intervene in the world. Hence, these tools generate information \textit{for} reality.
    \item Finally, AI as a \textbf{sociotechnical imaginary} gives room for organisations to introduce interpretations of the capabilities of AI-based systems and the possible consequences of their use on reality, inviting individuals to make decisions based on these interpretations instead of on reality itself. Hence, such an interpretation becomes information \textit{as} reality.
\end{enumerate}

In the presented order, these modes of interaction between humans and information feature increasing difficulty in exerting virtuous behaviour, ensuring alignment with moral principles, and assessing the effects of actions upon reality. Hence, they also feature an increasing difficulty in assessing ethical quality of actions, following the framework outlined in Section \ref{ethics}.

Given the review of the AI landscape described in the present article, the actual implementation of the proposed framework to assess the ethical quality of actions using AI tools should be possible and helpful to ensure that AI tools are used for the benefit of society. As discussed in previous paragraphs, the implementation of this framework can be based on a first-order logical theory establishing virtue-driven ethical principles and corresponding methods for proper use and quality assessment of AI technology. An executable specification of the framework is under development and shall be presented in future publications.

\bibliographystyle{alpha}
\bibliography{sample}

\end{document}